\documentclass[aip,pof,reprint]{revtex4-1}

\usepackage{amsmath,amsfonts,amssymb}%
\usepackage{docs}%
\usepackage{bm}%
\usepackage{graphicx}%
\usepackage{subfig}%
\usepackage[colorlinks=true,linkcolor=blue]{hyperref}%

\newcommand{\argmin}{\operatornamewithlimits{argmin}}

\expandafter\ifx\csname package@font\endcsname\relax\else
 \expandafter\expandafter
 \expandafter\usepackage
 \expandafter\expandafter
 \expandafter{\csname package@font\endcsname}%
\fi
\begin{document}

\title{Towards Scalable Parallel-in-Time Turbulent Flow Simulations}%

\author{Qiqi Wang}%
\email[Corresponding author, ]{qiqi@mit.edu}
\homepage{engineer-chaos.blogspot.com}
\affiliation{Department of Aeronautics and Astronautics, MIT,
77 Massachusetts Ave, Cambridge, MA 02139, USA}%
\author{Steven A. Gomez}%
\email{gomezs@mit.edu}
\affiliation{Department of Aeronautics and Astronautics, MIT,
77 Massachusetts Ave, Cambridge, MA 02139, USA}%
\author{Patrick J. Blonigan}%
\email{blonigan@mit.edu}
\affiliation{Department of Aeronautics and Astronautics, MIT,
77 Massachusetts Ave, Cambridge, MA 02139, USA}%
\author{Alastair L. Gregory}%
\email{blonigan@mit.edu}
\affiliation{Department of Aeronautics and Astronautics, MIT,
77 Massachusetts Ave, Cambridge, MA 02139, USA}%
\author{Elizabeth Y. Qian}%
\email{blonigan@mit.edu}
\affiliation{Department of Aeronautics and Astronautics, MIT,
77 Massachusetts Ave, Cambridge, MA 02139, USA}%

\date{October 2012}%
\revised{May 2013}%

\begin{abstract}
We present a reformulation of unsteady turbulent flow simulations. The
initial condition is relaxed and information is allowed to propagate
both forward and backward in time. Simulations of chaotic dynamical
systems with this reformulation can be proven to be well-conditioned
time domain boundary value problems. The reformulation can enable
scalable parallel-in-time simulation of turbulent flows.
\end{abstract}

\maketitle


\section{Need for space-time parallelism}

The use of computational fluid dynamics (CFD) in science and engineering can
be categorized into \emph{Analysis} and \emph{Design}.  
A CFD \emph{Analysis} performs a
simulation on a set of manually picked parameter values.
The flow field is then inspected to gain understanding of the flow
physics.  Scientific and engineering decisions are then made based on
understanding of the flow field.  Analysis based on high fidelity
turbulent flow simulations, particular Large Eddy Simulations, is a
rapidly growing practice in complex engineering applications
\cite{0957-0233-12-10-707} \cite{lesCombustionRev}.

CFD based \emph{Design} goes beyond just performing individual simulations,
towards sensitivity analysis, optimization, control, uncertainty
quantification and data based inference.  \emph{Design} is enabled by
\emph{Analysis} capabilities, but often requires more rapid turnaround.
For example, an engineer designer or an
optimization software needs to perform a series of simulations,
modifying the geometry based on previous
simulation results.  Each simulation must
complete within at most a few hours in an industrial design environment.
Most current practices
of design use steady state CFD solvers, employing RANS (Reynolds
Averaged Navier-Stokes) models
for turbulent flows.  Design using high fidelity, unsteady
turbulent flow simulations has been investigated in academia\cite{MaWaDeMo07}.
Despite their great potential, high fidelity design is infeasible
in an industrial setting because each simulation typically takes days to
weeks.

\begin{figure}[htb!] \centering
\vspace{0.00\textwidth}
\includegraphics[width=0.48\textwidth]{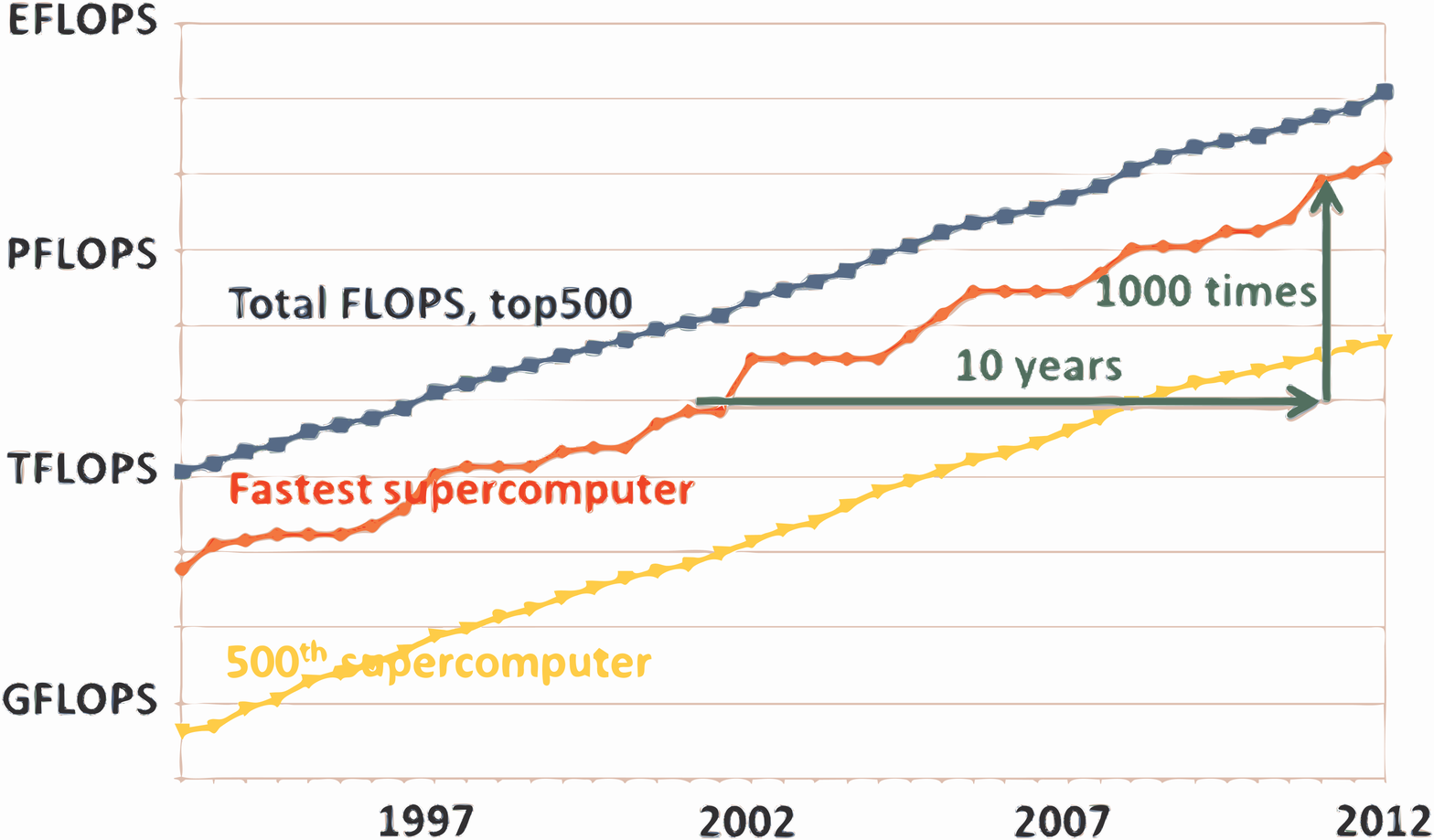}
\caption{Exponential increase of high performance computing power,
primarily sustained by increased parallelism in the past decade.
Data originate from top500.org. GFLOPS, TFLOPS, PFLOPS
and EFLOPS represent $10^{9}$, $10^{12}$, $10^{15}$ and $10^{18}$
FLoating point Operations Per Second, respectively.}
\vspace{0.00\textwidth}
\label{f:top500}
\end{figure}
The inability of performing high fidelity turbulent flow
simulations in short turnaround time is a barrier to the
game-changing technology of high fidelity CFD-based design.
Nevertheless, development in High Performance Computing (HPC), as
shown in Figure \ref{f:top500}, promises
to delivery in about ten years computing hardware a thousand times
faster than those available today.  This will be achieved through
extreme scale parallelization of light weight, communication
constrained cores \cite{springerlink10}.
A 2008 study developed a straw man extreme scale system that could deliver
$10^{18}$ FLOPS by combining about 166 million cores
\cite{kogge2008exascale} \cite{amarasinghe2009exascale}.
CFD simulations running on a million cores can
be as common in a decade as those running on a thousand cores today.

Will the projected thousand-fold increase in the number of computing cores
lead to a thousand-fold decrease in turnaround time of high fidelity turbulent
flow simulations?  If the answer is yes, then the same LES that takes a week
to complete on today's systems would only take about 10 minutes in 2022.
High fidelity CFD-based design in an industrial setting would then be a
reality.

\begin{figure}[htb!] \centering
\vspace{-0.00\textwidth}
\includegraphics[width=0.45\textwidth]{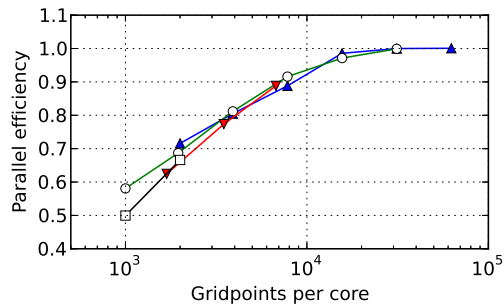}
\vspace{-0.00\textwidth}
\caption{Typical LES in complex geometry\cite{moin2002advances} suffers
from decreased parallel efficiency when the number of gridpoints per core
is less than 10,000.  The upward triangles, circles,
downward triangles and squares represent meshes of 2M, 1M, 216k and
64k grid points, respectively.}
\vspace{-0.00\textwidth}
\label{f:paraeff}
\end{figure}
The answer to this question hinges on development of more efficient, scalable
and resilient parallel simulation paradigm.  Current unsteady flow
solvers are typically parallel only in space.  Each core handles the
flow field in a spatial subdomain.  All cores advance simultaneously
in time, and data is transferred at subdomain boundaries at each time step.
This current simulation paradigm can achieve good parallel scaling when
the number of grid points per core is large.  However, Figure
\ref{f:paraeff} shows that parallel efficiency quickly deteriorates
as the number of grid points per core decreases below a few thousand.
This limit is due to limited inter-core
data transfer rate, a bottleneck expected to remain or worsen in the
next generation HPC hardware.  Therefore, it would be inefficient to run a
simulation with a few million grid points on a million next-generation
computing cores.  We need not only
more powerful computing hardware but also a next-generation simulation
paradigm in order to dramatically reduce the turnaround time of unsteady
turbulent flow simulations.

\begin{figure}[htb!] \centering
\includegraphics[width=0.48\textwidth]{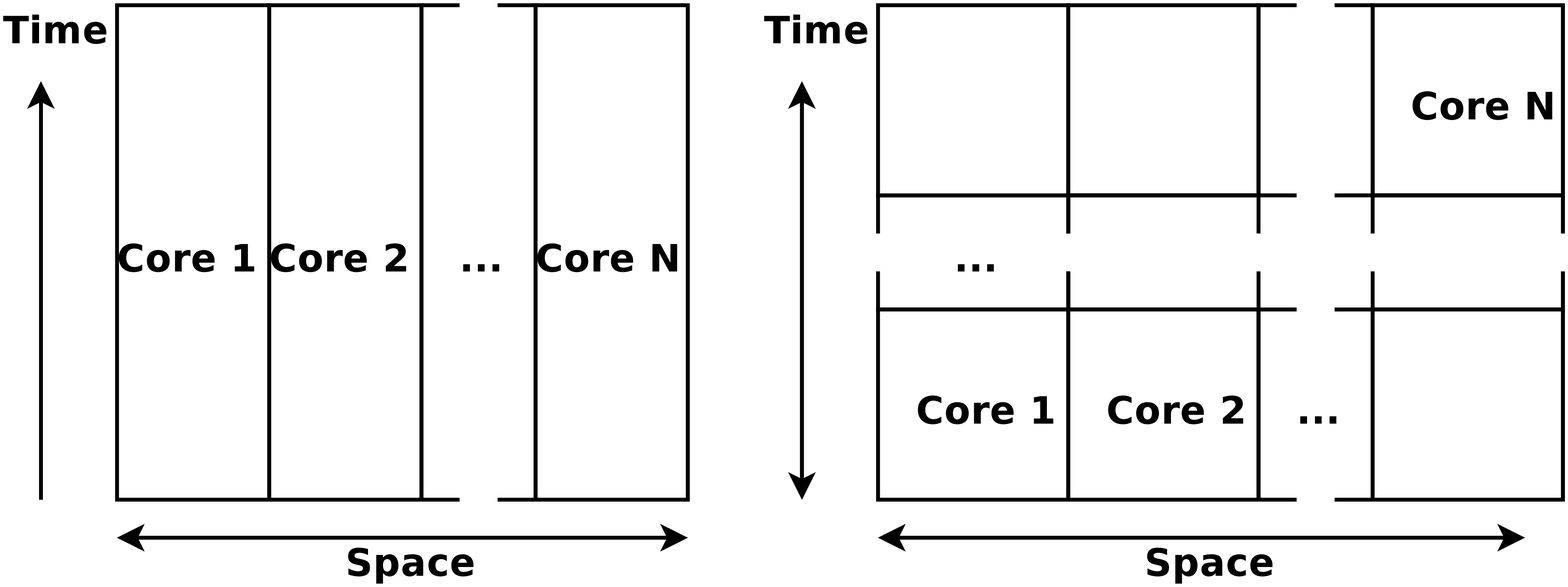}
\vspace{-0.00\textwidth}
\caption{Illustration of spatial parallelism (left) and space-time
parallelism (right).}
\vspace{-0.00\textwidth}
\label{f:spacetime}
\end{figure}

A key component of this enabling, next-generation simulation paradigm
can likely be \emph{space-time parallel simulations}.  These simulations
subdivide the 4-dimensional space-time computational domain.  
Each computing core handles a contiguous subdomain of the simulation
space-time.  Compared to subdivision only in the 3-dimensional space,
space-time parallel simulations can achieve significantly higher level
of concurrency, and reduce the ratio of inter-core communication to
floating point operations.  Each core
computes the solution over a fraction of the entire simulation time
window (Fig \ref{f:spacetime}), reducing the simulation turnaround time.

\emph{Space-time parallel simulations} can significantly reduce 
the ratio of inter-core data transfer to floating point operations.
This ratio can be estimated from the fraction of
grid points lying on subdomain interfaces.
The fraction of interfacial grid points in a spatially parallel simulation
is estimated to be
$\sim 6 \left(M/N\right)^{-\frac13}$,
where $M$ is the total number of grid points assumed to be uniformly
distributed in a cubical domain, and $N$ is the number
of computing cores.
The fraction of space-time grid points in a space-time parallel simulation
is estimated to be $\sim 8 \left(M\,T/N\right)^{-\frac14}$, where $T$ is
the total number of time steps.  Table \ref{t:compare} shows that
typical space-time
parallel simulations on a million cores have significantly smaller ratio
of inter-core data transfer to floating point operations than equivalent
spatially parallel simulations.
\begin{table}[htb!]
\centering
\vspace{0.00\textwidth}
\begin{tabular}{|c|c|c|c|c|}
\hline
$M$ & $T$ & $m_{so}$ & $m_{st}$ & $\tilde{m}_{so}$ \\
\hline
$10^6$ & $10^5$ & $1$ & $10^5$ & $2.4\times 10^3$ \\
$10^7$ & $2\times 10^5$ & $10$ & $2\times 10^6$ & $2.2\times 10^4$ \\
$10^8$ & $5\times 10^5$ & $100$ & $5\times 10^7$  & $2.5\times 10^5$ \\
$10^9$ & $10^6$ & $1000$ & $10^9$  & $2.4\times 10^6$ \\
\hline
\end{tabular}
\caption{For typical simulations of $M$ grid points and $T$ time steps
running on $N=10^6$ cores, this table estimates the grid points per core
of a spatially parallel simulation $m_{so} = M/N$, the space-time grid
points per core of a space-time parallel simulation $m_{st} = M\,T/N$,
and the equivalent grid points per core of a spatially parallel simulation
with the same fraction of interfacial grid points as the space-time
parallel simulation $6 \tilde{m}_{so}^{-\frac13} = 8 m_{st}^{-\frac14}$.
Space-time parallel simulations have significantly more effective grid
points per core, which can lead to increased parallel efficiency.
}
\vspace{-0.00\textwidth}
\label{t:compare}
\end{table}
If a space-time parallel simulation achieves the same parallel
efficiency of a spatially parallel simulation for the same fraction of
interfacial grid points, Table \ref{t:compare} and Fig \ref{f:paraeff}
suggest that a typical million-grid-point turbulent flow simulation can
run efficiently on a million cores with good parallel efficiency.

By reducing the ratio of inter-core communication to floating point
operations, \emph{space-time parallel} simulations has the potential of
achieving high parallel efficiency, even for relatively small turbulent flow
simulations on extreme scale parallel machines.
Combined with next generation computing hardware, this could lead to
typical simulation turnaround time of minutes.  Space-time parallelism
could enable high fidelity CFD-based design, including sensitivity analysis,
optimization, control, uncertainty quantification and data-based inference.

\section{Barrier to efficient time parallelism}


Time domain decomposition methods have a long history
\cite{nievergelt1964parallel}.  Popular
methods include the multiple shooting method, time-parallel and
space-time multigrid method, and the Parareal method.
\begin{figure}[htb!] \centering
\vspace{-0.00\textwidth}
\includegraphics[width=0.48\textwidth]{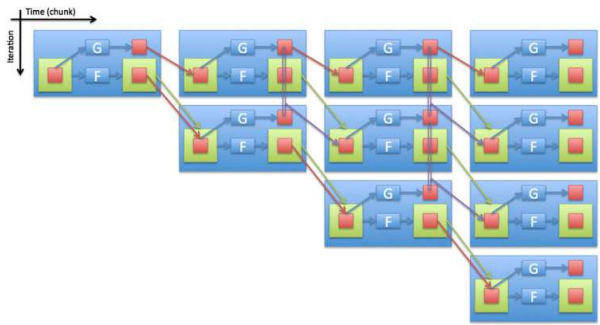}
\vspace{-0.00\textwidth}
\caption{Illustration of the Parareal algorithm, extracted from
reference\cite{reynolds2012mechanisms}.
$G$ and $F$ are the coarse and fine solvers, respectively.}
\vspace{-0.00\textwidth}
\label{f:parareal}
\end{figure}
As exemplified in Fig. \ref{f:parareal},
most time domain parallel methods divides the simulation time interval
into small time chunks.  They start with an initial estimate obtained
by a coarse solver.  Iterations are then performed over the entire
solution history, aiming to converge to the
solution of the initial value problem. In particular, the
\emph{Parareal} method
has been demonstrated to converge for large scale turbulent plasma
simulations\cite{Samaddar:2010:PTN:1831747.1831799, reynolds2012mechanisms}.

However, many time parallel methods suffer from a common scalability
barrier in the presence of chaotic dynamics.  The number of required
iterations increases as the length of the time domain increases.
As demonstrated for both the Lorenz attractor\cite{gander2008nonlinear}
and a turbulent plasma simulation\cite{reynolds2012mechanisms},
the number of iterations for reaching a given
tolerance is often proportional to the length of the time domain.
Because the the number of operations per iteration is also proportional to
the length of the time domain, the overall computation cost of most classical
algorithms scales with the square of the time domain length.
It is worth noting that recent developments have demonstrated a sub-quadratic
cost scaling with time domain length via processor reuse\cite{Berry20125945}.

\begin{figure}[htb!] \centering
\vspace{-0.00\textwidth}
\subfloat[$t=2$]{\includegraphics[width=0.4\textwidth]{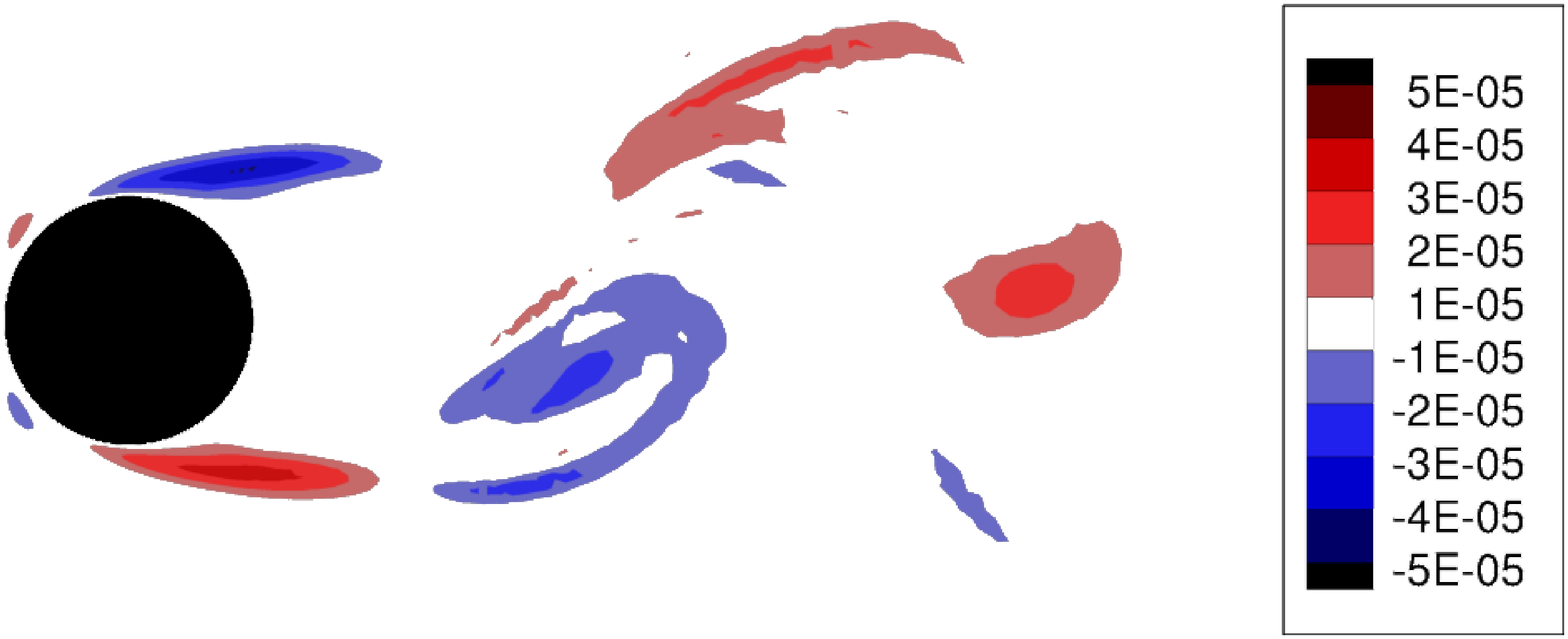}}
\subfloat[$t=22$]{\includegraphics[width=0.4\textwidth]{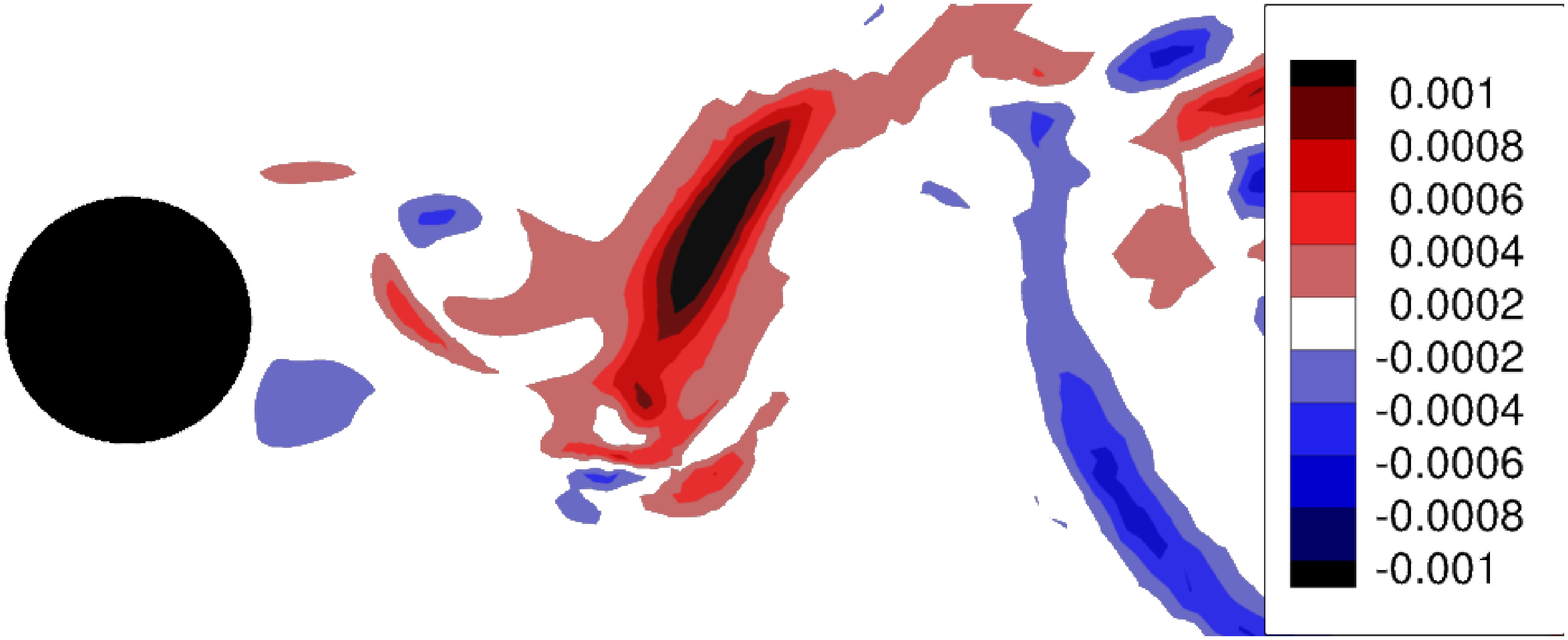}}\\
\subfloat[$t=62$]{\includegraphics[width=0.4\textwidth]{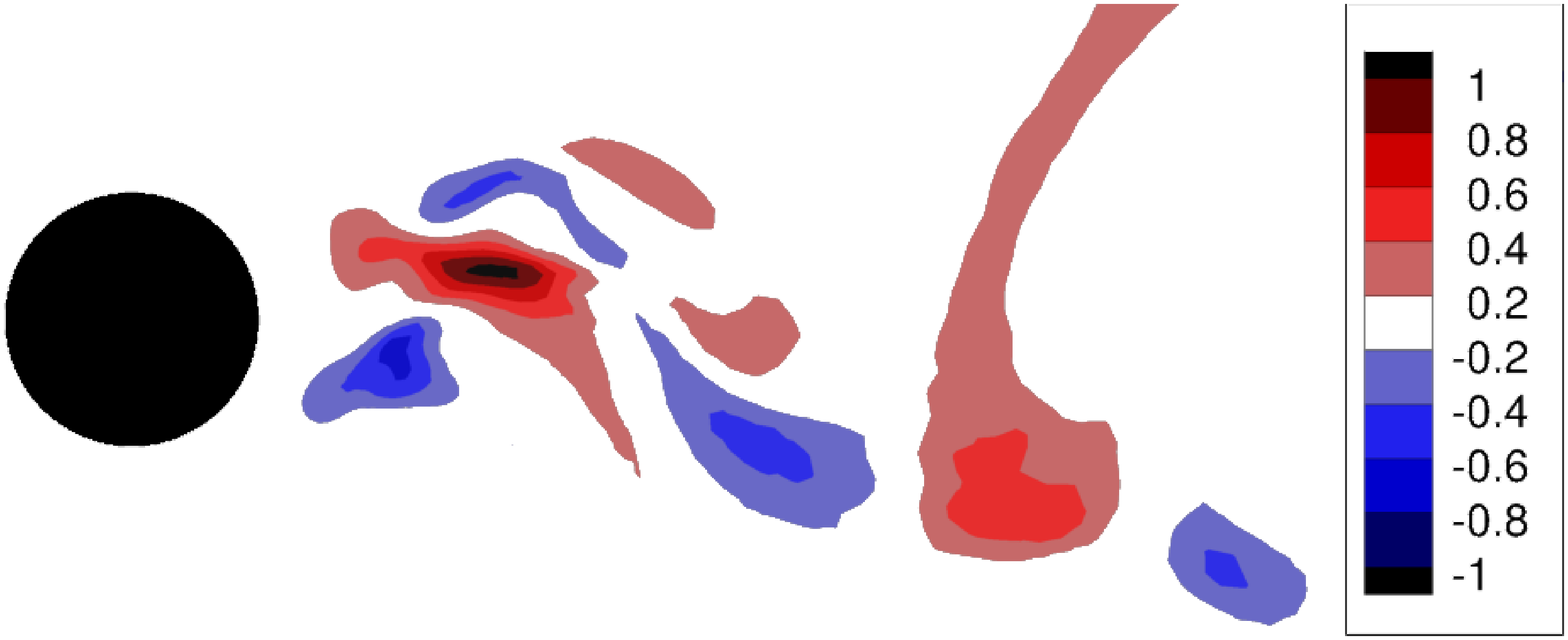}}
\vspace{-0.00\textwidth}
\caption{Spanwise velocity difference between two unsteady flow solutions
at $t=2,22$ and $62$.  A $10^{-5}$ magnitude perturbation at $t=0$
is the only difference between the two solutions.  The growing magnitude
of difference shows the ill-conditioning of a chaotic initial value
problem.  $Re_D=500$; periodic spanwise extent of $4D$ is used.}
\label{f:chaosRe500}
\vspace{-0.00\textwidth}
\end{figure}

This poor scalability is related to the characteristic sensitivity of chaos.
A small perturbation to a chaotic dynamical system, like the
turbulent flow shown in Fig \ref{f:chaosRe500}, can cause a large
change in the state of the system at a later time.
\begin{figure}[htb!] \centering
\includegraphics[width=0.49\textwidth]{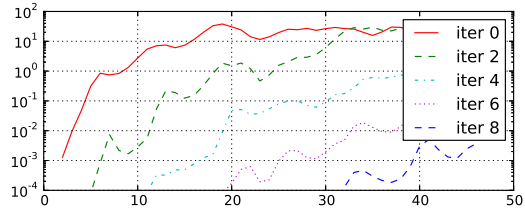}
\vspace{-0.00\textwidth}
\caption{Error in 9 Parareal iterations of
the Lorenz attractor\cite{gander2008nonlinear} shows increasingly delayed
convergence of later time chunks.  Horizontal axis represent
time chunks; each time chunk has length of 0.1.}
\vspace{-0.00\textwidth}
\label{f:lorenzParaReal}
\end{figure}
\begin{figure}[htb!] \centering
\vspace{0.00\textwidth}
\includegraphics[width=0.49\textwidth]{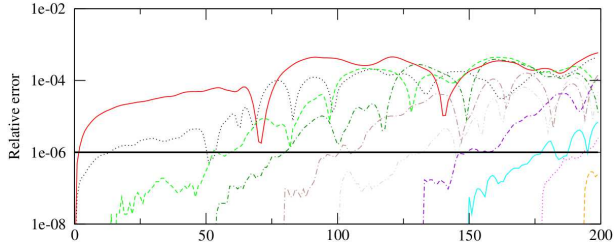}
\vspace{-0.00\textwidth}
\caption{Error in 12 Parareal iterations of a
turbulent plasma simulation shows increasingly delayed
convergence of later time chunks.  Figure is extracted from
reference\cite{reynolds2012mechanisms}.  Horizontal axis represent
time chunks.  The solid red line represents
the first iteration; the yellow dashed line at the lower right corner
represents the 12th iteration.}
\vspace{-0.00\textwidth}
\label{f:plasmaParaReal}
\end{figure}
This sensitivity causes a significant barrier
to fast convergence of time domain parallel methods.  In the Parareal
method, for example, a small difference between the coarse and
fine solvers in the early time chunks can cause a large difference in
the initial estimate and the converged solution.
A small correction made in the earlier time chunks can
result in a large update in later time chunks.  As a result, the later
time chunks can only converge after the earlier time chunks converge
(Figs \ref{f:lorenzParaReal} and \ref{f:plasmaParaReal}).  The number of
iterations required to converge the entire solution below a certain
tolerance therefore increases as the length of the time domain increases.

The cause of this poor scalability, the sensitivity of chaos,
can be quantified by the \emph{Lyapunov exponent}.
Consider two otherwise identical
simulations with an infinitesimal difference in their initial condition.
If the simulated system is chaotic, then the difference between these
two simulations would grow as $\exp \lambda t$.  This $\lambda$ is the
\emph{maximal Lyapunov exponent}, often just called as the
\emph{Lyapunov exponent}.  Mathematically, for a dynamical system
with an evolution function $\Phi^t(u)$,
\[ \lambda = \lim_{t\rightarrow\infty} \lim_{\epsilon \rightarrow 0}
   \frac1t \log \frac{\| \Phi^t(u+\epsilon v) - \Phi^t(u) \|}{\|\epsilon v\|}
\]
almost surely for any $v$ and $u$ on the attractor \cite{strogatzChaos}.

By describing how much a small
perturbation changes the solution at a later time,
the Lyapunov exponent $\lambda$ determines the convergence behavior of
time domain parallel methods.  A small error of magnitude $\epsilon$ at $t=0$
in a Parareal coarse solver
can cause an error of size $\sim\epsilon\: e^{\lambda t}$ at
a later time $t$.  A small update of size
$\epsilon$ at an earlier time $t_1$ can require an update of size
$\sim\epsilon\: e^{\lambda(t_2-t_1)}$ at a later time $t_1$.
It is no surprise that a larger Lyapunov exponent $\lambda$ poses a
greater challenge to time parallel methods like Parareal.

The influence of the Lyapunov exponent $\lambda$ on the convergence
of Parareal is particularly significant in chaotic dynamical systems
that have multiple time scales.  The maximal Lyapunov
exponent, with a unit of time$^{-1}$, is often inversely proportional
to the smallest time scale.  Consequently, time domain parallel methods
are particularly challenged by chaotic multiscale simulations in which
very small time scales are resolved.

\begin{figure}[htb!] \centering
\vspace{-0.00\textwidth}
\includegraphics[width=0.48\textwidth]{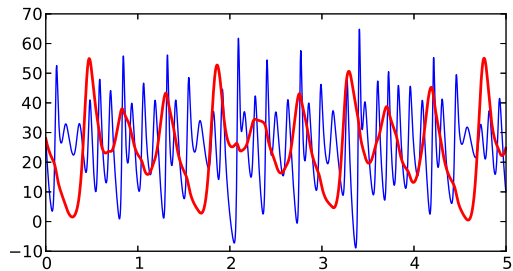}
\vspace{-0.00\textwidth}
\caption{A solution of the multiscale coupled Lorenz system
(\ref{coupledLorenz}) with $c=1/4$.  The thick, red line indicates
the slow variable $z_s$. The thin, blue line indicates the fast
variable $z_f$.}
\vspace{-0.00\textwidth}
\label{f:coupledTraj}
\end{figure}
\begin{figure}[htb!] \centering
\vspace{-0.00\textwidth}
\includegraphics[width=0.48\textwidth]{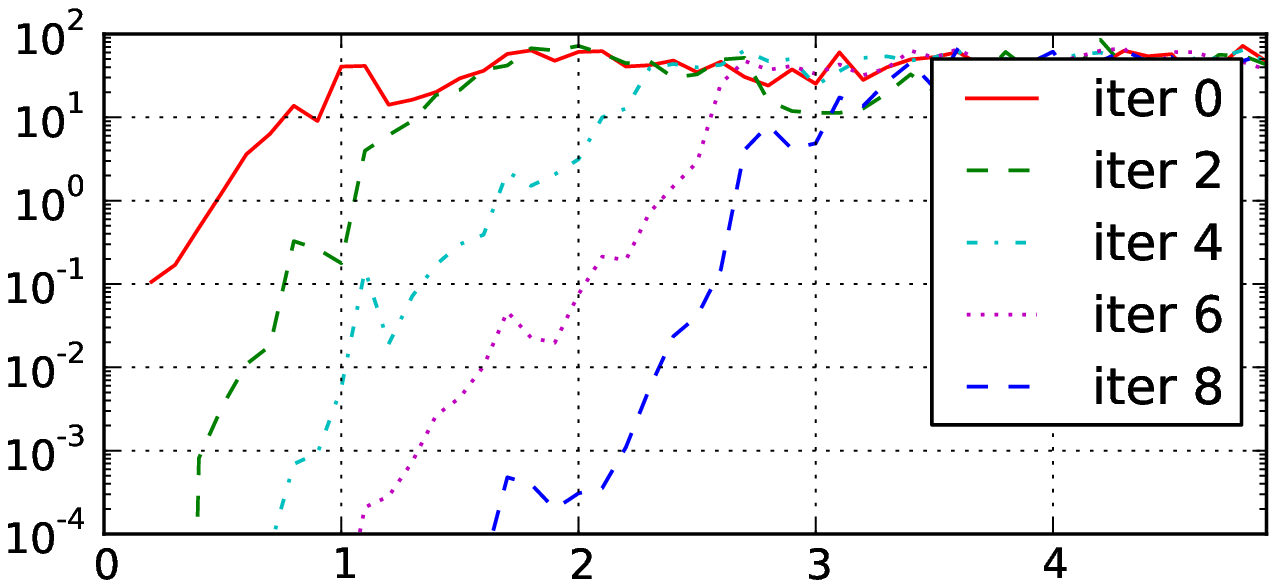}
\includegraphics[width=0.48\textwidth]{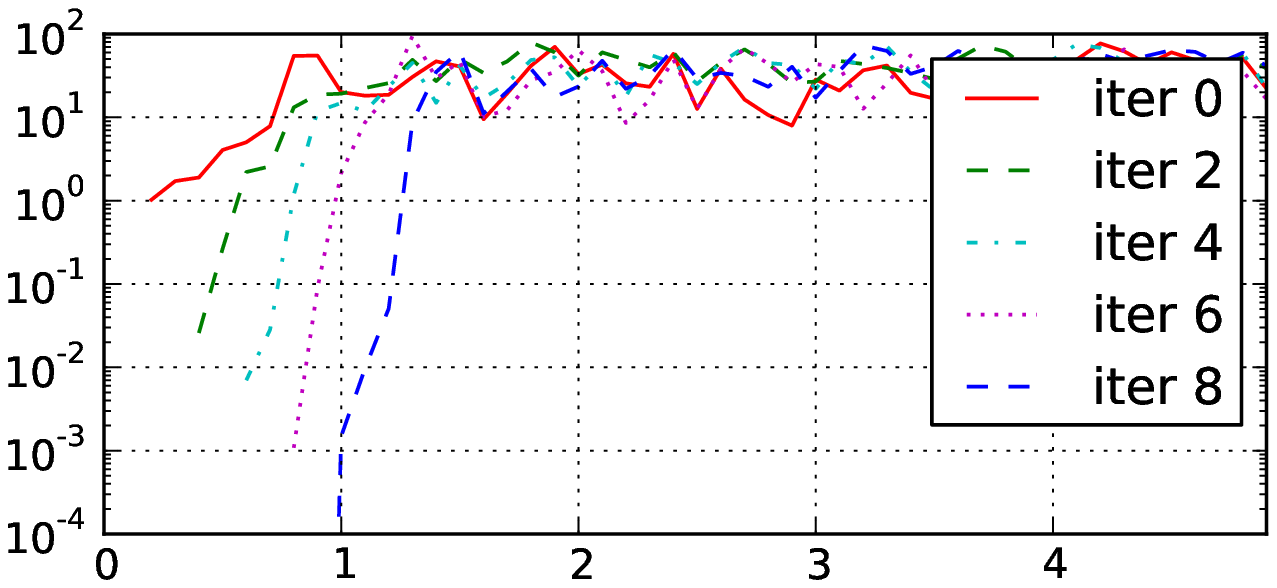}
\vspace{-0.00\textwidth}
\caption{Error in 9 Parareal iterations of the coupled Lorenz
system (\ref{coupledLorenz}) for $c=1/2$ (upper) and $c=1/4$ (lower).
We observe slower convergence when a wider range of chaotic timescales are
present.  The horizontal axes represent
time; each time chunk has length of 0.1.  The coarse solver is forward
Euler of step size $0.004\times c$; the fine solver is fifth order Runge Kutta
with adaptive time stepping.}
\vspace{-0.00\textwidth}
\label{f:coupledParareal}
\end{figure}
A simple example of such chaotic systems with multiple time scales is the
coupled Lorenz system
\begin{equation} \label{coupledLorenz}
\begin{aligned}
\dot{x}_s &= \sigma(y_s-x_s), & c\,\dot{x}_f &= \sigma(y_f-x_f) \\
\dot{y}_s &= x_s(r-z_s) - y_s,& c\,\dot{y}_f &= x_f(r-z_f) - y_f \\
\dot{z}_s &= x_s y_s - \beta (z_s+z_f),&
c\,\dot{z}_f &= x_f y_f - \beta (z_s+z_f)
\end{aligned}\end{equation}
Fig. \ref{f:coupledTraj} shows that
the coefficient $c<1$ determines the time scale of the fast
dynamics, whereas the slower dynamics has a time scale of about 1.
When $c$ decreases, the maximal Lyapunov exponent $\lambda$ increases.
Figure \ref{f:coupledParareal} shows that Parareal converges
proportionally slower for smaller $c$.

These pieces of evidence suggest that time domain parallel
methods such as Parareal can suffer from lack of scalability in
simulating chaotic multiscale systems, e.g., turbulent flows.
The length of the time domain in a turbulent flow simulation is often
multiples of the slowest time scale, so that converged statistics can be
obtained.  The finest resolved time scale that determines the maximal
Lyapunov exponent can be orders of magnitude smaller than the time
domain length.  Consequently, the required number of time domain parallel
iterations can be large.  Increasing the time domain length or the
resolved time scales would further increase the required number of iterations.

\section{Reformulating turbulent flow simulation}

Time domain parallel methods, such as Parareal, scale poorly
in simulation of chaotic, multiscale dynamical systems, e.g. turbulent
flows.  This poor scalability is because the {\bf initial value problem} of a
chaotic dynamical system is {\bf ill-conditioned}.  A perturbation of magnitude
$\epsilon$ at the beginning of the simulation can cause a difference
of $\epsilon\: \exp \lambda t$ at time $t$ later, where $\lambda$ is the
maximal Lyapunov exponent.  The condition number
of the initial value problem can be estimated to be
$$\kappa\sim\exp \lambda T\sim\exp \frac{T}{\tau}\;,$$
where $T$ is the time domain length and $\tau$ is the smallest resolved
chaotic time scale.  This condition number can be astronomically large for
multiscale simulations such as DNS and LES of turbulent flows.
Efficient time domain parallelism can only be achieved through
reformulating turbulent flow simulation into a well-conditioned problem.

We reformulate turbulent flow simulation into a well-conditioned problem
by relaxing the initial condition.  Not strictly enforcing an initial
condition is justified under the following two assumptions:
\begin{enumerate}
\item {\bf Interest in statistics}: all quantities of interest in
the simulation are statistics of the flow field in quasi-equilibrium
steady state.  These include the
mean, variance, correlation, high order moments and distributions
of the flow field.
\item {\bf Ergodicity}: starting from any initial condition, 
the flow will reach the same quasi-equilibrium steady state after initial
transient.  All statistics of the flow field are independent
of the initial condition after reaching the quasi-equilibrium steady state.
\end{enumerate}
Under these assumptions, satisfying a particular initial condition is
not important to computing the quantities of interest.  Instead of
trying to find {\bf the} flow solution that satisfies both the governing
equation and the initial condition, we aim to find {\bf a} flow solution
satisfying only the governing equation.
For example, we can formulate the problem as finding the solution of the
governing equation that is closest (in $L^2$ sense) to a reference
solution, which can come from a coarse solver or from solution at a
different parameter value.

Note that we do not address dynamic measures such as time domain
correlations and power spectra.  Further analysis is required to assess
whether these dynamic measures are captured by the present method.

Relaxing the initial condition annihilates the ill-conditioning in simulating
chaotic dynamical systems, making time parallelization
efficient and scalable.  If the initial condition were fixed, a
small perturbation near the beginning of the simulation would cause 
a large change in the solution later on.  If the initial condition were
relaxed, a small perturbation near the beginning
of the simulation could be accommodated by a small change in the initial
condition, with little effect on the the solution after the
perturbation.

\begin{figure}[htb!] \centering
\vspace{-0.00\textwidth}
\includegraphics[width=0.48\textwidth]{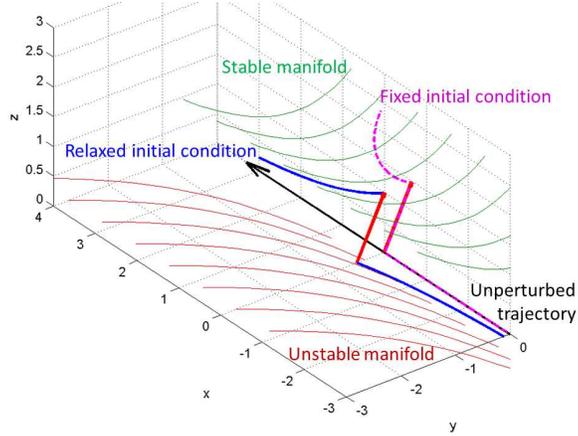}
\vspace{-0.00\textwidth}
\caption{A geometric illustration of the response of a chaotic dynamical
system to an impulse perturbation.  The black arrow represents an
unperturbed trajectory running along the $z$-axis.  The red arrows
represent the perturbation.  The magenta dashed line represents
the perturbed trajectory with fixed initial condition.  The blue line
represents a perturbed trajectory with relaxed initial condition.}
\vspace{-0.00\textwidth}
\label{f:response}
\end{figure}
The following geometric analysis demonstrates the
stability of a chaotic dynamical system with relaxed initial
condition.  The phase space around a trajectory can be decomposed into
a stable manifold and an unstable manifold.  Consider an arbitrary
impulse perturbation to the system as shown in Fig \ref{f:response}.
Because the perturbation can contain a component along the unstable
manifold, the perturbed trajectory would exponentially diverge from the
unperturbed trajectory if the initial condition were fixed.  However,
If the initial condition is relaxed, the perturbation along the
unstable manifold can be annihilated by slightly adjusting the initial
condition, as shown in Fig \ref{f:response}.  This small adjustment in the
initial condition makes it sufficient to only consider the
decaying effect of the perturbation along the stable manifold.  The effect of
the perturbation along the unstable manifold is reflected by the small
trajectory change before the perturbation.

Stability of the trajectory with a relaxed initial condition is achieved
by splitting a perturbation into stable and unstable components,
and propagating their effects forward and backward in time, respectively.
This stability property is formally known as the \emph{Shadowing Lemma}
\cite{pilyugin1999shadowing}.  Consider a dynamical system governed
by $$\frac{du}{dt} = \mathcal{R}(u)\;,$$
where $\mathcal{R}$ is a nonlinear spatial operator.
The shadowing lemma states that \emph{
For any $\delta>0$ there exists $\epsilon>0$, such that for every
$u_r$ that satisfies
$\| \partial u_r / \partial \tau - \mathcal{R} (u_r) \|
< \epsilon\,,\; 0\le t\le T \,,$
there exists a true solution $u$ and a time transformation $t(\tau)$,
such that $\|u(\tau) - u_r(\tau)\| < \delta$, $|1-dt/d\tau| < \delta$, and
$\partial u/\partial t - \mathcal{R} (u) = 0\,,\; 0\le \tau\le T $}.

\begin{figure}
\subfloat[Solutions of initial value problems]
{\includegraphics[width=0.45\textwidth]{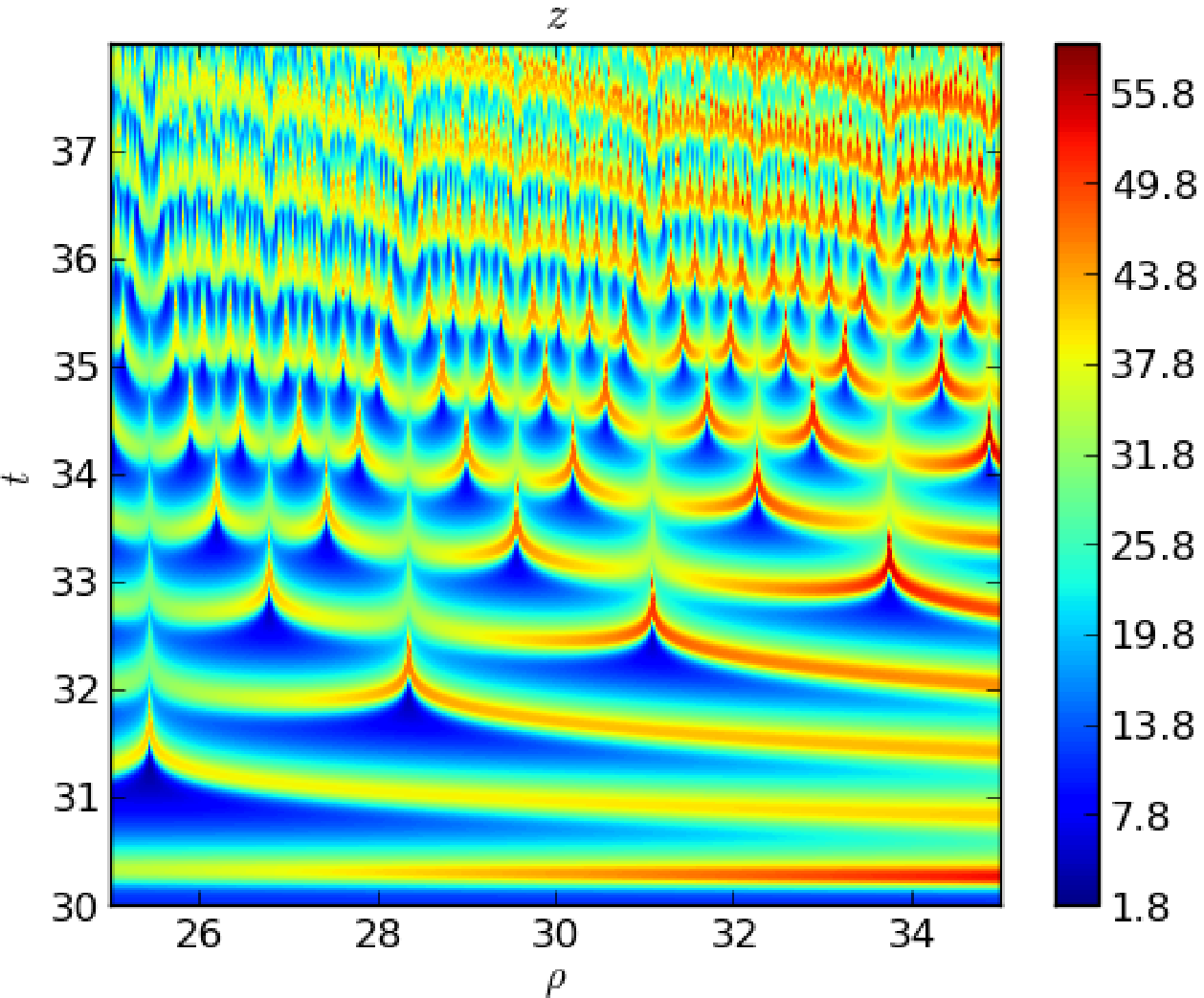}}
\hspace{0.02\textwidth}
\subfloat[Solutions of the reformulated system]
{\includegraphics[width=0.45\textwidth]{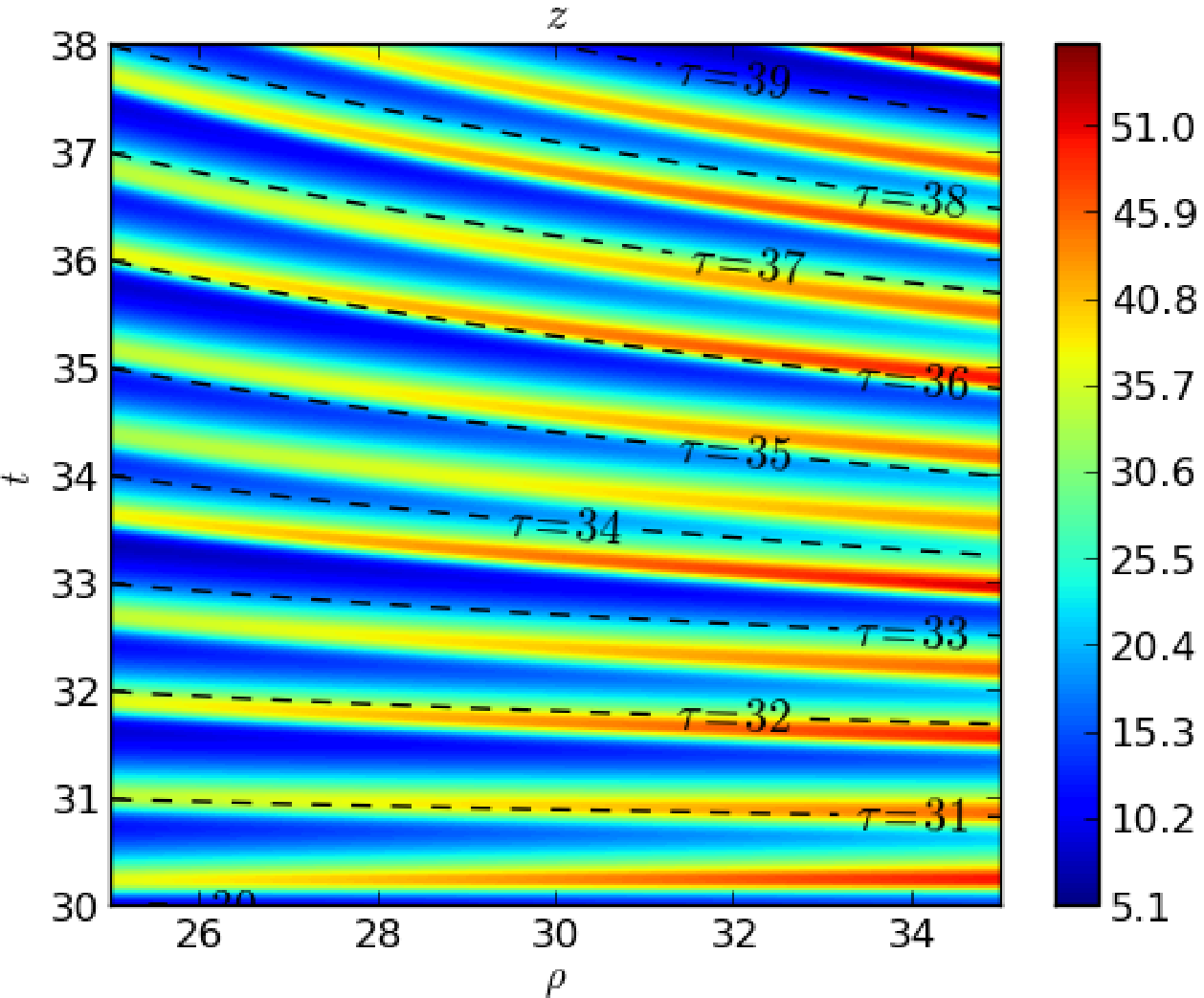}}
\caption{Comparison between solutions of initial value problems of the
Lorenz system (left, showing extreme sensitivity to $\rho$) and the
reformulated system (\ref{lsqnonlinear}) (right, showing smooth
dependence on $\rho$).  Note that the solutions of the reformulated
system at different values of $\rho$ ``shadow'' each other.
The difference in the physical time $t$ and
the shadowing time $\tau$ is the effect of nonzero time dilation
$\eta(\tau)$.}
\label{f:ivplsp}
\end{figure}

The shadowing lemma theorizes the trajectory stability of chaotic
dynamical systems with relaxed initial conditions.
This paper develops the following least squares formulation designed
to take advantage of this stability numerically
\begin{equation} \label{lsqnonlinear} \begin{split}
    u,\eta &= \argmin \frac12 \int_0^T \Big(\| u(\tau)-u_{r}(\tau) \|^2
                                     + \eta(\tau)^2\Big) d\tau\\
    &\mbox{s.t.}\quad
      (1+\eta)\frac{\partial u}{\partial\tau}
    = \mathcal{R}(u)\;,\quad 0\le \tau\le T
\end{split}\end{equation}
This least squares problem
finds a solution $u(\tau)$ satisfying the differential equation
$$ \frac{\partial u(\tau(t))}{\partial t} = \mathcal{R}(u)$$ 
with a time transformation $\tau$ satisfying $d\tau/dt = 1+\eta(\tau)$.
Here the scalar function of time $\eta(\tau)$ is part of the solution to
the least squares problem.  The solution $u(\tau), \eta(\tau)$
minimizes a combination of two metrics, (a) the distance between
$u(\tau)$ and a reference solution $u_{r}(\tau)$, and (b) the deviation from
unity of the time transformation, represented by $\eta(\tau)$.

The least squares problem (\ref{lsqnonlinear}) has a unique and stable
solution when the reference solution $u_{r}$ is close to the actual
solution $u$, so that $u - u_{r}$ can be described by the corresponding
linearized least squares problem.
With a convex, positive definite objective function and a
linear constraint, this problem has a unique and stable solution.

The constraint least squares problem (\ref{lsqnonlinear}) can be simplified
by forming its Lagrange function
\begin{equation} \begin{split}
\Lambda = \int_0^T &\bigg(
             \frac{\langle u-u_{r}, u-u_{r}\rangle+\eta^2}{2}
               + \Big\langle w, \frac{\partial u}{\partial t}
               - \mathcal{R}(u)\Big\rangle\bigg)\, dt
\end{split} \end{equation}
where $w$ is the Lagrange multiplier.
Variation of the Lagrange function can be transformed through integration by
parts
\begin{equation} \label{nonlinearibp} \begin{split}
\delta\Lambda =& \int_0^T
    \delta\eta\Big(\eta + \Big\langle w, \frac{\partial u}{\partial\tau}
              \Big\rangle\Big)
               + \langle \delta u, u-u_{r}\rangle 
               + \Big\langle w, \frac{\partial\delta
               u}{\partial t} - \mathcal{L}\delta u\Big\rangle\, dt\\
           =& \int_0^T \Big\langle \delta u\;,\; u-u_{r}
             - \frac{\partial w}{\partial t}
             - \mathcal{L}^* w \Big\rangle\,dt 
            + \int_0^T\delta\eta\;\Big(\eta + \Big\langle w\;,\;
               \frac{\partial u}{\partial\tau} \Big\rangle\Big)\,dt
             + \langle \delta u, w\rangle \bigg|_0^T
\end{split} \end{equation}
where $\mathcal{L}$ is the linearized operator
\begin{equation}
\mathcal{L} = \left.\frac{\delta \mathcal{R} (u)}{\delta u}\right|_{u_r}\;,
\end{equation}
$\mathcal{L}^*$ is the adjoint operator of $\mathcal{L}$.
The solution of the constraint least squares problem
(\ref{lsqnonlinear}) satisfies the first order optimality condition
$\delta\Lambda=0$ for all $\delta u$ and $\delta\eta$.
Therefore, Equation (\ref{nonlinearibp}) leads to
\begin{equation} \label{uandw}
u = u_r + \frac{\partial w}{\partial t} + \mathcal{L}^* w\;,
\quad \eta = -\Big\langle w\;,\;
               \frac{\partial u}{\partial\tau} \Big\rangle
\end{equation}
and the boundary condition $w(\tau=0) = w(\tau=T) = 0$.

Substituting these equalities back into the constraint in
(\ref{lsqnonlinear}), we obtain the following second order
boundary value problem in time:
\begin{equation} \label{nonlinearbvp} \begin{split}
  & \frac{\partial}{\partial t}
    \left(\frac{\partial w}{\partial t} + \mathcal{L}^* w + u_r\right)
  - \mathcal{R}\left(\frac{\partial w}{\partial t}
  + \mathcal{L}^* w + u_r\right) = 0 \\
  & w\big|_{\tau=0} = w\big|_{\tau=T} = 0\;,
\end{split} \end{equation}
where the transformed time derivative is
\begin{equation}
 \frac{\partial}{\partial t} = (1+\eta)\,\frac{\partial}{\partial \tau}\;.
\end{equation}

Our reformulation of chaotic simulation produces a least squares
system (\ref{lsqnonlinear}), whose solution $u$ satisfies the
governing equation, and has relaxed initial condition.  This solution
can be found by first solving the Lagrange multiplier $w$ through
Equation (\ref{nonlinearbvp}), a second order boundary value problem in
time, then computing $u$ from Equation (\ref{uandw}).
As illustrated in Figure \ref{f:ivplsp},
this reformulated problem does not suffer from ill-conditioning
found in initial value problems of chaos, therefore overcomes
a fundamental barrier to efficient time-parallelism.
The reformulated boundary value
problem in time (\ref{nonlinearbvp}) is well-conditioned and suitable
for time domain parallel solution methods.  These properties are demonstrated
mathematically in Section \ref{s:algnonlinear} and Appendix \ref{s:app},
and through examples in Sections \ref{s:lorenz} and \ref{s:ks}.

\section{Solving the reformulated system}
\label{s:algnonlinear}

Solving the reformulated problem involves two steps: 1. solving a nonlinear
boundary-value-problem in time (\ref{nonlinearbvp}) for the Lagrange
multiplier $w$, and 2. evaluating Equation (\ref{uandw}) for the solution $u$.
Performing the second step in space-time parallel is relatively
straightforward.  This section introduces our iterative algorithm for
the first step based on Newton's method.
\begin{enumerate}
\item Decompose the 4D spatial-temporal computational domain into
subdomains.
\item Start at iteration number $k=1$ with an initial guess 
$u^{(0)}$ distributed among computing cores.
\item Compute the residual $du^{(k-1)}/dt - \mathcal{R}(u^{(k-1)})$
in each subdomain.
Terminate if the magnitude of the residual meets the convergence
criteria.
\item {\bf Solve a linearized version of Equation (\ref{nonlinearbvp})
for $w^{(k)}$ using a space-time parallel iterative solver} (detailed
follows in the rest of this section).
\item Compute the time dilation factor $\eta$ and transformed time
$dt = d\tau\,\exp\,(-\eta(\tau))$.
\item Compute the updated solution $u^{(k)}$ by parallel evaluation of
Equation (\ref{uandw}) with $u_r = u^{(k-1)}$ and $w=w^{(k)}$.
\item Update to transformed time by letting $\tau=t$; continue to Step 3
with $k=k+1$.
\end{enumerate}
Note that the update in Step 5 avoids degeneracy when $\eta\le-1$, and is
first order consistent with $dt = d\tau/(1+\eta(\tau))$ implied
by the definition of $\eta$.  Numerically, this step is involves
updating each physical time step size $\Delta t_i$ from the corresponding
shadowing time step size $\Delta\tau_i$, while keeping the number of time
steps fixed.

The only step that requires detailed further explanation is Step 4.  All
other steps can be evaluated in space-time parallel with minimal
communication across domain boundaries.
Solving Equation (\ref{nonlinearbvp}) for $w$ using Newton's
method requires a linear approximation to the equation.
Ignoring second order terms by assuming both $u-u_r$ and $\eta$ are
small, Equation (\ref{nonlinearbvp}) becomes
\begin{equation} \label{linearbvp} \begin{split}
  -& \frac{\partial^2 w}{\partial \tau^2}
  - \left(\frac{\partial}{\partial \tau}\mathcal{L}^*
        - \mathcal{L}\frac{\partial}{\partial \tau} \right) w
  + \big(\mathcal{L}\mathcal{L}^* + \mathcal{P}\big) w
  = f_{\tau} \\
  & w\big|_{\tau=0} = w\big|_{\tau=T} = 0\;,
\end{split} \end{equation}
where $\mathcal{P}$ is a symmetric positive semi-definite spatial operator
\begin{equation} \label{defP}
   \mathcal{P}w = \left\langle w\,,\;
   \frac{\partial u_r}{\partial t}\right\rangle
   \frac{\partial u_r}{\partial t}\;.
\end{equation}
and
\begin{equation}
f_{\tau} = \frac{\partial u_r}{\partial\tau} - \mathcal{R}(u_r)
\end{equation}
It can be shown that the system (\ref{linearbvp}) is a symmetric
positive definite system with good condition number (see Appendix \ref{s:app}).
It also has a energy minimization form, allowing a standard finite
element discretization.  Appendix \ref{s:disc} details the
discretization and the numerics of solving Equation (\ref{linearbvp}).

We implemented both a direct solver and an
iterative solver for the symmetric boundary value problem (\ref{linearbvp}).
The computation time of both methods is analyzed in 
Appendix \ref{s:disc}.  The total cost of a direct parallel solution method
(cyclic reduction) is estimated to be 10 times that of solving a linear
initial value problem of the same size using implicit time stepping and Gauss
elimination in each time step.  If a parallel iterative solution method
(multigrid-in-time
\cite{brandt1981multigrid}) is used, the total computation cost of solving
Equation (\ref{linearbvp}) with $n_{iter}$ total smoothing iterations
is about 6 times that of solving a linear initial value
problem of the same size, with $n_{iter}$ iterations performed in
solving each implicit time step.  Our naive implementation of
multigrid-in-time requires an $n_{iter}$ ranging from 100 to 1000 for
the Lorenz system and the Kuramoto-Sivashinsky equation.  This number could be
reduced through more research into iterative solution algorithm
specifically for Equation (\ref{linearbvp}).
When the operator $\mathcal{L}$ is a large
scale spatial operator, we envision that the multigrid-in-time method can
be extended to a space-time multigrid method.  

\section{Validation on the Lorenz System}
\label{s:lorenz}

The algorithm described in Section \ref{s:algnonlinear} is applied
to the Lorenz system.  Both the solution $u(t)$ and the Lagrange
multiplier $w(t)$ in this case
are in the Euclidean space $R^3$.  The three components of the solution
are denoted as $u(t) = (x, y, z)$ by convention.
The governing equation is
\begin{equation}
\dot{u} = \mathcal{R}(u) = \Big(s(y-x),\; x(r-z) - y,\; x y - b z \Big)\;,
\end{equation}
where $s=10$, $r=35$ (the Rayleigh number) and $b=8/3$ are three parameters
of the system.

\begin{figure}[htb!] \centering
\vspace{0.00\textwidth}
\includegraphics[width=0.45\textwidth]{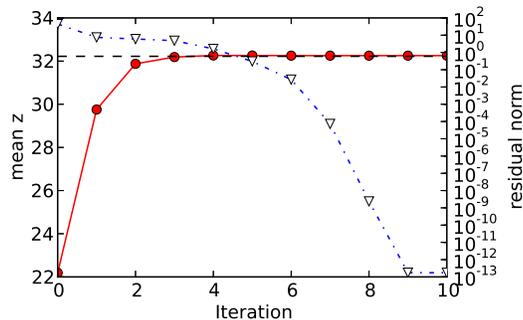}
\caption{Solution of the Lorenz system at $s,r,b=10,{\bf 35},8/3$
with the algorithm of Section \ref{s:algnonlinear}.
The initial guess $u^{(0)}(t), t\in[0,100]$ is a solution of the system at
$s,r,b=10,{\bf 25},8/3$.
The horizontal axis indicates the iteration number $k$.
The dot-dash line with open triangles indicates the
norm of the residual $f_{\tau}^{(k)}$.
The red, solid line with filled circles represents
the time averaged $z$ of the solution $u^{(k)}$.  The horizontal dashed
line indicates the mean $z$ computed from an independent initial value
solution at $r={\bf 35}$.}
\vspace{0.00\textwidth}
\label{f:lorenzconverge}
\end{figure}

The initial guess $u^{(0)}$ used in Section
\ref{s:algnonlinear}'s algorithm is a numerical solution of the
system at a different Rayleigh number $r=25$.  The initial guess has 10,001
time steps uniformly distributed in a time domain of length 100.
Appendix \ref{s:disc} discusses the details of the numerics.  

\begin{figure}[htb!] \centering
\vspace{0.00\textwidth}
\includegraphics[width=0.45\textwidth]{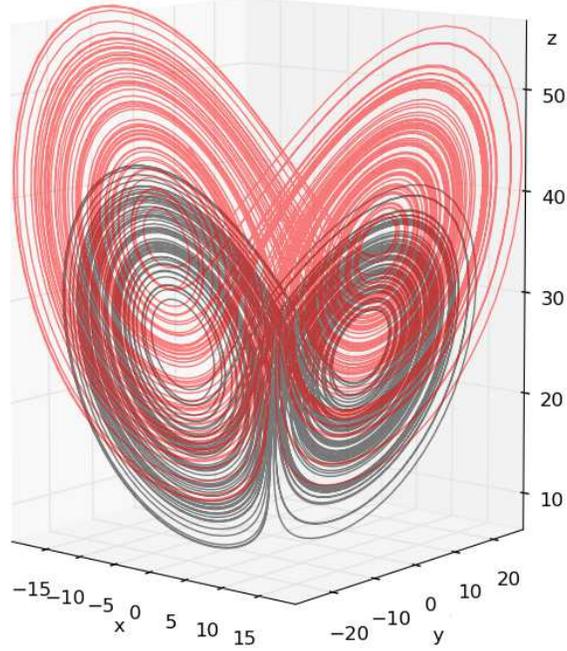}
\caption{The black trajectory is a phase space plot of the initial guess
$u^{(0)}$, a solution of the Lorenz system at $s,r,b=10,{\bf 25},8/3$.
The red trajectory plots the converged solution $u^{(10)}$, a least
squares solution of the Lorenz system at $s,r,b=10,{\bf 35},8/3$.}
\vspace{0.00\textwidth}
\label{f:attractor}
\end{figure}

\begin{figure}[htb!] \centering
\vspace{0.00\textwidth}
\includegraphics[width=0.48\textwidth]{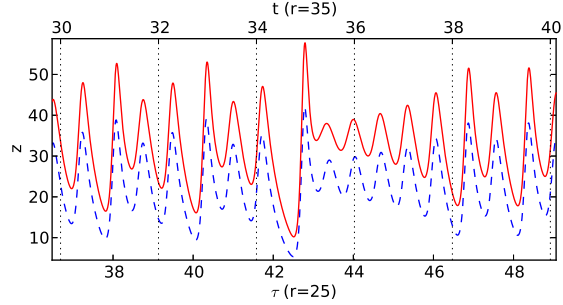}
\caption{The blue, dashed line represents $z(\tau)$ of the initial guess,
a solution of the system at $r=25$.
The red, solid line represents $z(t)$ of the converged solution at
$r=35$.}
\vspace{0.00\textwidth}
\label{f:shadow}
\end{figure}

As a general rule in Newton's method for solving nonlinear systems,
the algorithm described in Section \ref{s:algnonlinear} 
is expected to converge only when the initial guess is sufficiently
close to the solution.  In fact, we rarely observe convergence to a
physical solution if one starts from an arbitrary
constant initial guess.  However, we do observe convergence when the
initial guess is obtained from a solver with a much large time step size,
or when the initial guess is a qualitatively similar solution at a
different parameter value.  The example shown in Figures
\ref{f:lorenzconverge}, \ref{f:attractor} and \ref{f:shadow} has a
challenging initial guess at a very different parameter value ($\rho=25$
versus $35$.)  The final solution has an averaged magnitude of over $50\%$
large than the initial guess.

Figure \ref{f:lorenzconverge} shows that the Newton's iterations
described in Section
\ref{s:algnonlinear} converges to machine precision within 9 iterations.
Less iterations are required when the initial
guess is closer to the final solution.
The phase space plots in Figure \ref{f:attractor}
shows that the converged trajectory lies on the attractor of the
Lorenz system at $r=35$, and is significantly different from the initial
guess trajectory.  Figure \ref{f:shadow} shows that the converged
trajectory closely tracks the initial guess trajectory on a transformed time
scale.
The statistical quantity (mean $z$) of the converged solution
matches that computed from a conventional time integration.

\begin{figure}[htb!] \centering
\vspace{0.00\textwidth}
\includegraphics[width=0.36\textwidth]{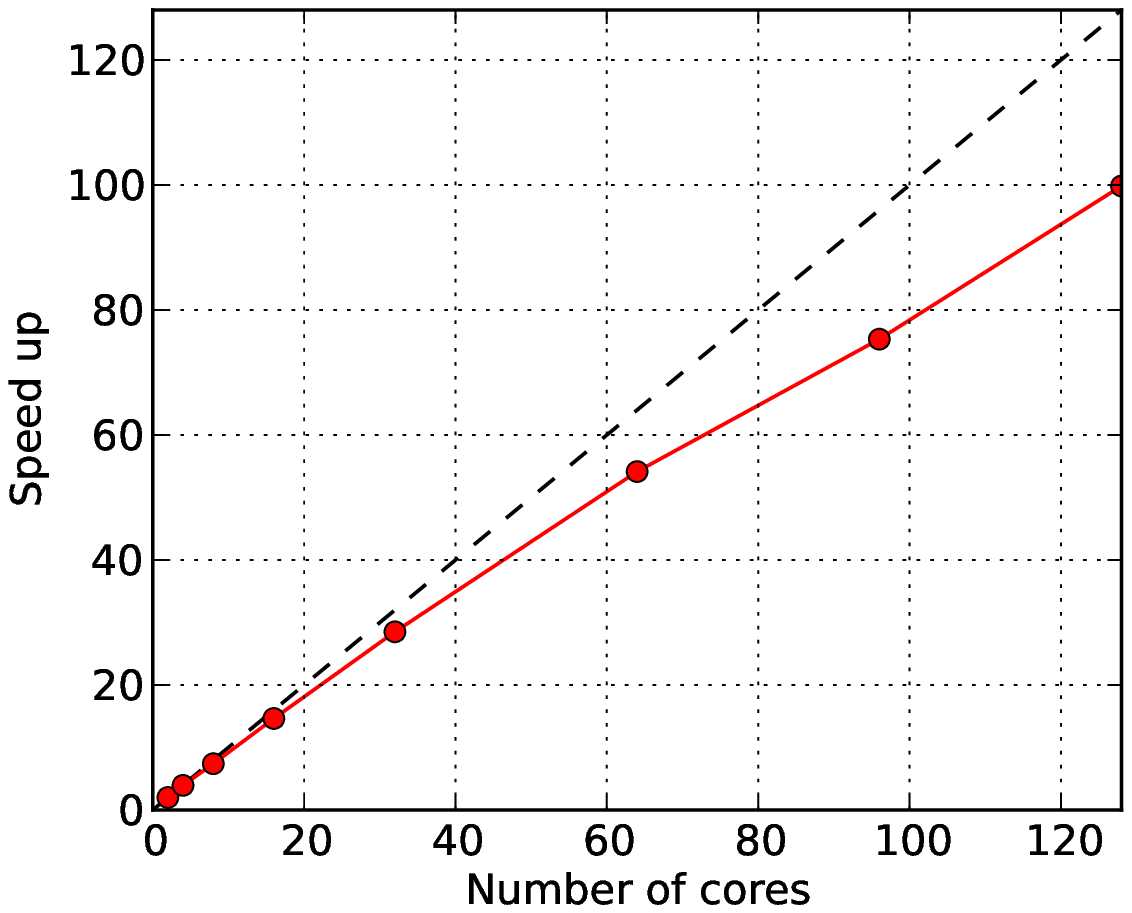}
\caption{Speed up of the multigrid-in-time solver of the linearized
system (\ref{linearbvp}) using up to 128 cores.  4096 time steps
are allocated per core with uniform time step size of 0.01.  The same
convergence criterion is used for all cases.}
\vspace{0.00\textwidth}
\label{f:speedup}
\end{figure}
Because the reformulated system do not suffer from the ill-conditioning of
the initial value problem,
parallelization in time domain does not encounter the same problem as
Parareal method for solving initial value problems.  Figure
\ref{f:speedup} shows that our algorithm is scalable up to a
time domains of length over 5000.  In comparison, Parareal would take
many more iterations to converge on longer time domains.
Converging on a time domain of length 5000 would require over 500
Parareal iterations based on estimates from Figure \ref{f:lorenzParaReal}.

\section{Validation on the KS Equation}
\label{s:ks}

The algorithm described in Section \ref{s:algnonlinear} is applied
to the Kuramoto-Sivashinsky equation.
Both the solution $u(x,t)$ and the Lagrange
multiplier $w(x,t)$ in this case are functions in the spatial domain
of $x\in[0, 100]$.  They also both satisfy the boundary conditions
\[ u\Big|_{x=0,100} = \frac{\partial u}{\partial x}\bigg|_{x=0,100}
 = 0 \]
The governing equation is defined by
\begin{equation} \label{ks}
\frac{\partial u}{\partial t} = 
\mathcal{R}(u) = -(u + c) \frac{\partial u}{\partial x}
- \frac{\partial^2 u}{\partial x^2}
- \frac{\partial^4 u}{\partial x^4}\;,
\end{equation}

\begin{figure}[htb!] \centering
\vspace{0.00\textwidth}
\includegraphics[width=0.46\textwidth]{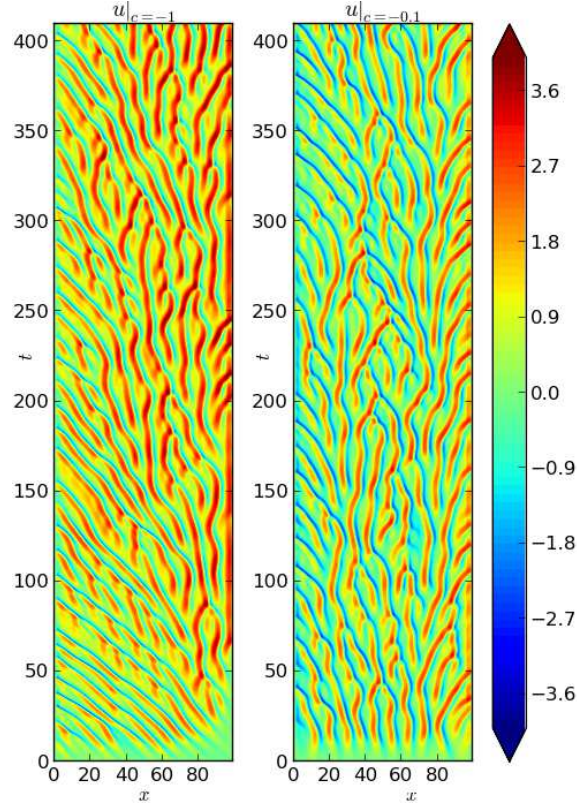}
\vspace{0.00\textwidth}
\caption{Initial value solution of the Kuramoto-Sivashinsky equation
(\ref{ks}) for $c=-1$ (left) and $c=-0.1$ (right).  Finite difference
discretization on 128 uniform spatial gridpoints is used.  The initial
condition in both cases use i.i.d. uniform $[-\frac12, \frac12]$ random
numbers.}
\label{f:ksivp}
\end{figure}
where the parameter $c$ is introduced to control the behavior of the system.
Figure \ref{f:ksivp} shows that solutions of this initial boundary value
problem at $c=-0.1$ and $c=-1$ reach different chaotic quasi-equilibrium
after starting from similar initial conditions.  The mean
of the solution at equilibrium is higher in the $c=-1$ case than in the
$c=-0.1$ case.

\begin{figure}[htb!] \centering
\vspace{0.00\textwidth}
\includegraphics[width=0.46\textwidth]{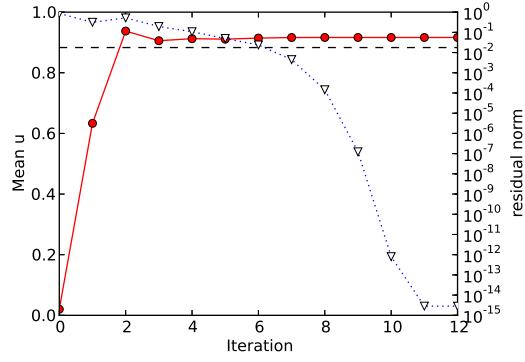}
\vspace{0.00\textwidth}
\caption{Solution of the Kuramoto-Sivashinsky (KS) equation (\ref{ks}) at $c=-1$
with the algorithm of Section \ref{s:algnonlinear}.
The initial guess $u^{(0)}(x,t), x\in[0,100], t\in[0,50]$
is a solution of the KS equation at $c=-0.1$.
The horizontal axis indicates the iteration number $k$.
The dot-dash line with open triangles indicates the
norm of the residual $f_{\tau}^{(k)}$.
The red, solid line with filled circles represents
the solution $u^{(k)}(x,t)$ averaged over $x$ and $t$.  The horizontal dashed
line indicates the averaged $u(x,t)$ computed from an independent initial
value problem at $c=-1$.}
\label{f:ksconverge}
\end{figure}

Section \ref{s:algnonlinear}'s algorithm is tested for the
Kuramoto-Sivashinsky equation at $c=-1$.
The initial guess $u^{(0)}$ is a numerical solution of the
Kuramoto-Sivashinsky equation at $c=-0.1$.  This initial guess has 401
time steps uniformly distributed in a time domain of length 100.
Appendix \ref{s:disc} discusses the details of the numerics.  

\begin{figure}[htb!] \centering
\vspace{0.00\textwidth}
\includegraphics[width=0.36\textwidth]{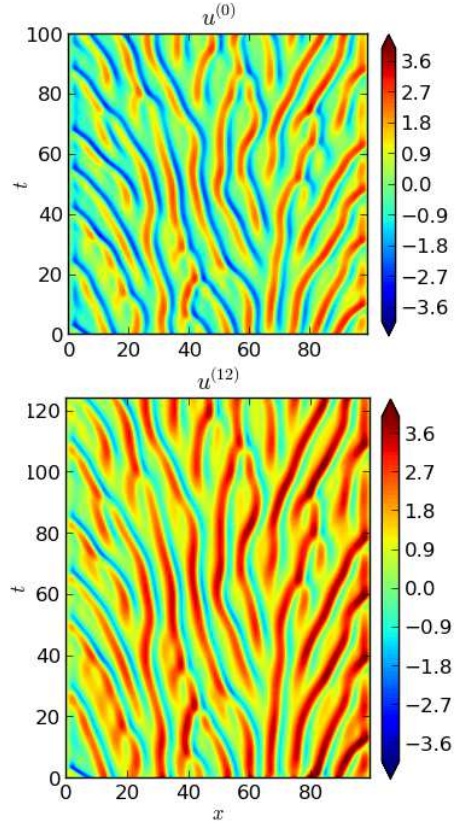}
\vspace{0.00\textwidth}
\caption{The initial guess $u^{(0)}$ and the converged solution
$u^{(12)}$ in Section \ref{s:algnonlinear}'s algorithm applied to the
Kuramoto-Sivashinsky equation (\ref{ks}) at $c=-1$.}
\label{f:ksshadow}
\end{figure}

The Newton's iteration converges even though the initial guess has an
advection velocity $c$ that is 10 times lower than the solution.
Figure \ref{f:ksconverge} shows that the algorithm of Section
\ref{s:algnonlinear} converges on the Kuramoto-Sivashinsky equation
to machine precision within about 10 iterations.
The statistical quantity (mean $u$) of the converged solution is close
to that computed from a time integration.  The small discrepancy between them
may be caused by the difference between the infinite time statistical average
and the finite time average over the time domain length of about 120.
Figure \ref{f:ksshadow} shows that the converged
solution shadows the initial condition on a stretched time scale.

\section{Conclusion}
We used the ergodic hypothesis to relax the initial condition in
simulating chaotic, unsteady dynamical systems.  The system with relaxed
initial condition do not suffer from ill-conditioning encountered in
fixed-initial condition problems.  Consequently, efficient parallel-in-time
simulation can be performed without scalability problems
from which time parallelization of initial value problem suffers.
The relaxed-initial-condition system is formulated into a least squares
problem, whose solution can be obtained through a second order boundary
value problem in time.  We solve this boundary value problem using an
iterative solution algorithm based on Newton's method.  All steps in
the algorithm can be efficiently performed in time-parallel.

This methodology is demonstrated on simulations of the Lorenz system and
the Kuramoto-Sivashinsky equation.  The iterative algorithm converges
in both cases to machine precision solutions
within 10 iterations.  The statistical quantities of
the converged solutions with relaxed initial conditions match those
computed from traditional time integration methods.  Time parallel
scalability of this method is demonstrated on the Lorenz attractor.

In summary, the primary advantages of this methodology are
\begin{enumerate}
\item Scalable time-parallelism.  The reformulation can enable a new class of
time-parallel and space-time-parallel computational simulation codes
that effectively use next generation exascale computers.
\item Well-conditioning.  The proposed formulation can be proved
to have low condition numbers, making many simulated-based computational
methods, e.g., optimization, uncertainty quantification and inference
easier to apply.
\end{enumerate}
The associated disadvantages of this methodology are mainly
\begin{enumerate}
\item Increased number of floating point operations.  The total number of
floating point operations is estimated to be 6 to 10 times that of
solving an initial value problem of the same size using implicit time
stepping.  The ratio can potentially be higer if a larger number of
Newton iterations or linear solver iterations are required.
\item Requiring a initial guess.  We observe that the initial guess must be
qualitatively similar to a physical solution in order for the Newton's method to
converge.
\item Requiring research and developement of new solvers.  Research is
need to investigate iterative linear solvers that are efficient for
the large scale linear system involved in this methodology.  Existing
solvers based on initial value problems have to be rewritten to solve
the reformulated system.
\end{enumerate}

\section{Acknowledgments}
The authors thank financial support from NASA Award NNX12AJ75A through
Dr. Harold Atkins, AFOSR STTR contract FA9550-12-C-0065
through Dr. Fariba Farhoo, and a subcontract of DOE's Stanford PSAAP
to MIT.  The coauthors were supported by the ANSYS fellowship at MIT
Aerospace Computing and Design Lab and the NASA graduate summer
internship at Langley during the work.

\appendix
\section{Well conditioned}
\label{s:app}

The linearized equation (\ref{linearbvp}) also has a weak form
\begin{equation} \label{weak}
{\mathfrak a}(\delta w, w) + {\mathfrak l}(\delta w) = 0
\end{equation}
and an energy minimization form, which is the Lagrangian dual of the linearized
least squares (\ref{lsqnonlinear}),
\begin{equation} \label{amin}
w = \argmin_{w|_{0,T}=0} \left(\frac{{\mathfrak a}(w, w)}2
  + {\mathfrak l}(w)\right)\;,
\end{equation}
The symmetric, positive definite and well-conditioned bilinear form
$\mathfrak a$ is
\begin{equation} \label{bilinear}
   {\mathfrak a}(w,v)
   = \int_0^T\left\langle
     \frac{\partial w}{\partial\tau} + \mathcal{L}^* w,
     \frac{\partial v}{\partial\tau} + \mathcal{L}^* v\right\rangle
   + \left\langle w, \mathcal{P} v\right\rangle\, d\tau
\end{equation}
and linear functional $\mathfrak l$ is
\begin{equation}
  {\mathfrak l}(w) = \int_0^T\big\langle w, f_{\tau} \big\rangle \, d\tau\;.
\end{equation}
The derivation and properties of the weak and energy forms are detailed
in Appendix \ref{s:app}.
These forms of Equation (\ref{linearbvp}) makes it possible to achieve provable
convergence of iterative solution techniques.

By taking the inner product of a test function $v$ with both sides of Equation
(\ref{linearbvp}), integrating over the time domain $[0,T]$, and using
integration by parts, we obtain
\begin{equation}\label{energy}
     \int_0^T\left\langle
     \frac{\partial w}{\partial\tau} + \mathcal{L}^* w,
     \frac{\partial v}{\partial\tau} + \mathcal{L}^* v\right\rangle
   + \left\langle w, \mathcal{P} v\right\rangle
   - \big\langle v, f_{\tau} \big\rangle\, d\tau = 0\;,
\end{equation}
which is the weak from (\ref{weak}).
We decompose the solution $w$ into the adjoint Lyapunov eigenvectors
in order to show that the symmetric bilinear form $\mathfrak{a}(u,v)$
as in Equation (\ref{bilinear}) is positive
definite and well-conditioned.  The Lyapunov decomposition
for an $n$-dimensional dynamical systems (including discretized PDEs) is
\begin{equation}\label{lyap}
w(t) = \sum_{i=1}^n w_i(t) \phi_i(t) \;,\quad\mbox{where}\quad
\frac{\partial\phi_i}{\partial t} + \mathcal{L}^*\phi_i = \lambda_i \phi_i
\end{equation}
Each $w_i$ is a real valued function of $t$, and $\phi_i(t)$ is the
corresponding adjoint Lyapunov eigenvectors.
By convention, the Lyapunov exponents are sorted such that
$\lambda_i \ge \lambda_{i+1}$.
Combining the Lyapunov eigenvector decomposition (\ref{lyap})
results in
\[ \frac{\partial w}{\partial\tau} + \mathcal{L}^* w
 = \sum_{i=1}^n \left( \frac{dw_i}{d\tau} + \lambda_i w_i\right) \phi_i\]
By substituting this equality and Equations (\ref{defP})
into the bilinear form and using the positive semi-definiteness of
$\mathcal{P}$, we obtain
\begin{equation} \label{aa} \begin{split}
    &\mathfrak{a}(w,w) \ge \\
  & \int_0^T 
     \sum_{i,j=1}^n\Big(\frac{dw_i}{d\tau} + \lambda_i w_i\Big)
               \Big(\frac{dw_j}{d\tau} + \lambda_j w_j\Big)
               \langle\phi_i,\phi_j\rangle\, d\tau
\end{split} \end{equation}
The eigenvalues of the matrix
$\Big(\langle\phi_i,\phi_j\rangle\Big)_{i,j=1,\ldots,n}$
are bounded away from both zero and infinity for uniformly hyperbolic
dynamical systems, i.e., there exists $0<c\le C<\infty$ such that
\begin{equation} \begin{split}
 c \sum_{i=1}^n x_i^2 <
 \sum_{i,j=1}^n x_i\,x_j \langle\phi_i,\phi_j\rangle < C \sum_{i=1}^n x_i^2
\end{split} \end{equation}
for any $x_1,\ldots,x_n$.  By applying these bounds to $\langle w,w\rangle$
and to Equation (\ref{aa}), we obtain
\begin{equation} \label{wavebound1} \begin{split}
\mathfrak{a}(w, w) &
> c \sum_{i=1}^n \int_0^T \left(\frac{dw_i}{d\tau} + \lambda_i
  w_i\right)^2\,d\tau\\ 
\langle w, w\rangle &
 < C \sum_{i=1}^n \int_0^T w_i^2\,d\tau
\end{split} \end{equation}
Most high dimensional, ergodic chaotic systems of practical interest are
often assumed to be quasi-hyperbolic, whose global properties are not
affected by their non-hyperbolic nature.  Therefore we conjecture that
(\ref{wavebound1}) also holds for these
quasi-hyperbolic systems when $T$ is sufficiently large.
Because of the boundary conditions $w(0) = w(T)=0$,
\begin{equation} \label{wavebound2} \begin{split}
& \int_0^T \left(\frac{dw_i}{d\tau} + \lambda_i w_i\right)^2\,d\tau \\
=& \int_0^T \left(\frac{dw_i}{d\tau}\right)^2 d\tau + \lambda_i
  \underbrace{\int_0^T w_i\frac{dw_i}{d\tau} \,d\tau}_{=0} + 
  \lambda_i^2 \underbrace{\int_0^T w_i^2 \,d\tau}_{\ge 0} \\
\ge & \int_0^T \left(\frac{dw_i}{d\tau}\right)^2 d\tau
\end{split}\end{equation}
Also because of the boundary conditions,
each $w_i(t)$ admits a Fourier sine series
\begin{equation}
    w_i(t) = \sum_{m=1}^{\infty} \hat{w}_{im} \sin \frac{m\pi t}{T} \,,
  = \sum_{m=1}^{\infty} \hat{w}_{im}\,\frac{m\pi }{T} \cos \frac{m\pi t}{T}\;.
\end{equation}
thus,
\begin{equation}
    \frac{dw_i}{d\tau}
  = \sum_{m=1}^{\infty} \hat{w}_{im}\,\frac{m\pi }{T} \cos \frac{m\pi t}{T}\;.
\end{equation}
The Parseval identity applies to both orthogonal series, leading to
the Poincare inequality
\begin{equation} \begin{split}
  & \int_0^T \left(\frac{dw_i}{d\tau}\right)^2 d\tau
  = \frac{T}2 \sum_{m=1}^{\infty} \hat{w}_{im}^2 
    \underbrace{\left(\frac{m\pi }{T}\right)^2}_{\ge \left(\pi/T\right)^2} \\
  \ge & \left(\frac{\pi}{T}\right)^2
      \frac{T}2 \sum_{m=1}^{\infty} \hat{w}_{im}^2 
  = \left(\frac{\pi}{T}\right)^2\int_0^T w_i^2\, d\tau
\end{split} \end{equation}
By combining this inequality with the inequalities (\ref{wavebound1})
and (\ref{wavebound2}), we obtain
\begin{equation}
\frac{\mathfrak{a}(w,w)}{\langle w,w\rangle} >
\left(\frac{\pi}{T}\right)^2 \frac{c}{C} 
\end{equation}
This inequality leads to our conclusion that the symmetric bilinear form
$\mathfrak{a}(w,v)$ is positive definite.  Equation (\ref{linearbvp})
is well-conditioned because if the weak form (\ref{weak})
holds and $\mathfrak{a}(w,w) = \mathfrak{l}(w)$, then
\begin{equation}
\|\mathfrak{l}\| \|w\| \ge \mathfrak{l}(w) = 
\mathfrak{a}(w,w) > \left(\frac{\pi}{T}\right)^2 \frac{c}{C}  \|w\|^2\;,
\end{equation}
therefore,
\begin{equation}
\frac{\|w\|}{\|\mathfrak{l}\|} < \left(\frac{\pi}{T}\right)^2 \frac{C}{c}\;.
\end{equation}
This inequality bounds the magnitude of the solution by the magnitude of the
perturbation.  It also bounds the magnitude of the solution error by
the magnitude of the residual.  This condition
number bound is $C/c$ times that of the Poisson equation in a 1D domain
$[0,T]$.

\section{Discretization and numerical solution}
\label{s:disc}

Equation (\ref{amin}) can be solved using the Ritz method, which is
equivalent to applying Galerkin projection on (\ref{linearbvp}).
The time domain $[0,T]$ is discretized into $n$ intervals by
$0=\tau_0 < \tau_1 < \ldots < \tau_{n} = T$.
Both the trial function $w$ and the
test function $v$ are continuous function that is piecewise linear
(i.e., linear within each interval $\tau_{i-1}\le \tau\le \tau_i$).
The operators $\mathcal{L}^*$ and $\mathcal{P}$ are approximated as
piecewise constant.  Denote $w_i = w(\tau=\tau_i), i=0,\ldots,n$,
$\mathcal{L}^*_i=\mathcal{L}^*(\tau_{i-1}<\tau<\tau_i)$ and
$\mathcal{P}_i=\mathcal{P}(\tau_{i-1}<\tau<\tau_i)$, Equation (\ref{bilinear})
can be written as
\[ {\mathfrak a}(w,v)
  = \Big(v_1^T\, v_2^T\, \ldots\,v_{n-1}^T\Big)
    \left(\begin{array}{cccc}
    A_1 & B_2 & & \\
    B_2^T & A_2 & \ddots & \\
       & \ddots & \ddots & B_{n-1} \\
       &        & B_{n-1}& A_{n-1} \end{array}\right)
    \left(\begin{array}{c}
    w_1\\ w_2 \\ \vdots \\w_{n-1}\end{array}\right)
\]
where
\[ A_i = \frac{I}{\Delta t_{i+1}} + \frac{I}{\Delta t_i}
       + \mathcal{L}_i + \mathcal{L}^*_i
       - \mathcal{L}_{i+1} - \mathcal{L}^*_{i+1}
       + \frac{\Delta t_i}{3} \Big(\mathcal{L}_i\mathcal{L}^*_i +
                                   \mathcal{P}_i\Big)
       + \frac{\Delta t_{i+1}}{3} \Big(\mathcal{L}_{i+1}\mathcal{L}^*_{i+1} +
                                   \mathcal{P}_{i+1}\Big) \] 
\[ B_i = -\frac{I}{\Delta t_i}
       + \mathcal{L}^*_i + \mathcal{L}_i
       + \frac{\Delta t_i}{6} \Big(\mathcal{L}_i\mathcal{L}^*_i +
                                   \mathcal{P}_i\Big) \] 
Therefore, Equation (\ref{linearbvp}) can be solved through a
symmetric, positive definite, block-tridiagonal system
\begin{equation} \label{tridiagonal} \left(\begin{array}{cccc}
    A_1 & B_2 & & \\
    B_2^T & A_2 & \ddots & \\
       & \ddots & \ddots & B_{n-1} \\
       &        & B_{n-1}& A_{n-1} \end{array}\right)
    \left(\begin{array}{c}
    w_1\\ w_2 \\ \vdots \\w_{n-1}\end{array}\right) = 
    \left(\begin{array}{c}
    l_1\\ l_2 \\ \vdots \\l_{n-1}\end{array}\right)
\end{equation}

This system (\ref{tridiagonal}) can be solved in parallel using either
direct or iterative method.  The estimated computational runtime of a
drect and an iterative parallel method are listed below.
Here $N$ is the number of domain decompositions in
the time domain. $M$ is the spatial degree of freedom, i.e., the size of
the blocks $A_i$ and $B_i$ in (\ref{tridiagonal}).
\begin{enumerate}
\item {\bf Parallel cyclic reduction}
is efficient when the $M$ is small and the spatial operator
$\mathcal{L}$ is dense.  Its estimated computation time is \cite{Seal2013273}
\[ (C_{in} + 6\, C_{mm}) M^3 \left(\frac{n}{N} + \log N\right)
 + 2 \beta M^2 \left(\log \frac{n}{N} + 2 \log n\right)\;, \]
where $C_{in}$, $C_{mm}$ are the amortized time per floating point
operation for matrix inversion and matrix-matrix multiplication,
respectively.  $\beta$ is the average time to transmit one floating
point number between any two processing elements across the network.

In particular, when a single processor is used, i.e., $N=1$, the
computation time is about a factor of 10 higher than solving a
$M$-dimensional linear initial value problem with $N$ time steps
using implicit time stepping, where the linear system in each time step
is solved using Gauss elimination.
\item {\bf V-cycle parallel multigrid} is efficient when $M$ is large
and the spatial operator $\mathcal{L}$ is sparse.  Assume that the
matrix representation of $\mathcal{L}$ has on average $n_{nz}$ nonzero
entries per row.  A multigrid cycle containing $n_s$ pre- and post-smoothing
Jacobi-like iterations has an estimated computation time of
\[ 6\, C_{mv} M n_{nz} n_s \frac{n}{N} + 2 \beta M n_s \log \frac{n}{N}\;, \]
where $C_{mv}$ is the amortized time per floating point operation for
sparse matrix-vector multiplication.  $\beta$ is the average time to
transmit one floating point number between any two processing elements
across the network.

In particular, when a single processor is used, i.e., $N=1$, the
computation time of $n_{cyc}$ multigrid cycles is similar to that of solving a
$M$-dimensional linear initial value problem with $N$ time steps
using implicit time stepping, where the linear system in each time step
is solved using iterative method involving $6\,n_{cyc}\, n_{mv}$ Jacobi-like
iterations.
\end{enumerate}

\end{document}